# *I*-centered vs *F*-centered orthorhombic symmetry and negative thermal expansion of the charge density wave of $EuAl_2Ga_2$

Harshit Agarwal[1a], Surya Rohith Kotla[1], Leila Noohinejad[2], Biplab Bag[3], Claudio Eisele[1], Sitaram Ramakrishnan[4], Martin Tolkiehn[2], Carsten Paulmann[2], Arumugam Thamizhavel[5b], Srinivasan Ramakrishnan[6c], Sander van Smaalen[1d]

[1]*Laboratory of Crystallography, University of Bayreuth, Bayreuth, 95444, Germany*

[2]*P24, PETRA III, Deutsches Elektronen-Synchrotron DESY, Hamburg, 22607, Germany*

[3]*Department of Physics, Amity Institute of Applied Sciences, Amity University Jharkhand, Ranchi, 835303, India*

[4]*I-HUB Quantum Technology Foundation, Indian Institute of Science Education and Research, Pune 411008, India*

[5]*Department of Condensed Matter Physics and Materials Science, TIFR, Mumbai, 400005, India*

[6]*Department of Physics, Indian Institute of Science Education and Research, Pune 411008, India*

**Abstract:**

Together with $EuGa_4$ and $EuAl_4$, $EuAl_2Ga_2$ belongs to the $BaAl_4$ structure type with space group symmetry I4/mmm. $EuAl_2Ga_2$ develops an incommensurate charge density wave (CDW) at temperatures below $T_{CDW}$ = 51 K. On the basis of temperature dependent single-crystal X-ray diffraction (SXRD) data, the incommensurately modulated CDW crystal structure of $EuAl_2Ga_2$ is determined to possess orthorhombic superspace symmetry *Immm(00γ)s00*. This symmetry is different from the orthorhombic *Fmmm* based symmetry of the CDW state of $EuAl_4$. Nevertheless, both symmetries *Immm(00γ)s00* and *Fmmm(00γ)s00* lead to the same conclusion, that the CDW is supported by the layers of Al1 type atoms, while the Eu and Al2 or Ga atoms are not directly involved in CDW formation. The different symmetries of the CDW states of $EuAl_4$ and $EuAl_2Ga_2$, as well as the observation of negative thermal expansion in the CDW state of $EuAl_2Ga_2$ might be explained by the effects of Ga substitution in the latter compound.

***Email:** [a] harshit.physics@gmail.com,

[b] thamizh@tifr.res.in,

[c] ramky07@gmail.com,

[d] smash@uni-bayreuth.de

## Introduction:

Quantum materials have played a significant role in condensed matter research for over a decade, delivering groundbreaking results [1]. Quantum materials refer to a class of materials that exhibit intriguing quantum mechanical properties at the nanoscale or atomic scale [2]. Due to the advent of coupled topological materials [3], our knowledge of emergent processes in quantum materials has improved significantly in recent years [4]. Intermetallic materials with a centrosymmetric square net structural motif having the $BaAl_4$ structure type [5] are ideal examples of quantum materials that exhibit superconductivity (SC), charge density wave (CDW) ordering, antiferromagnetism, and heavy fermionic behavior [6]. In Eu-based intermetallics, Eu has two types of valance states: $Eu^{2+}$ (magnetic) and $Eu^{3+}$(nonmagnetic) due to an unstable *4-f* shell. Most Eu-based compounds exist in the divalent $Eu^{2+}$ states and follow the RKKY interaction that causes the long-range magnetic ordering [7]. Similarly, $Gd^{3+}$ ($4f^7$, J= 7/2, L=0) based centrosymmetric intermetallic compounds follow RKKY interaction, giving rise to Fermi surface nesting [8].

A charge density wave (CDW) is a spatial modulation of electron charge density coupled to a modulation of the atomic positions (periodic lattice modulation, PLD) [9,10]. The origin of CDWs varies between materials but it can be classified according to the dimensionality of the band structure. In quasi-one-dimensional (1D) metals, the stabilization of CDW can be explained by the Peierls model, including Fermi surface nesting, metal-insulator transition, and periodic lattice distortion (PLD). In contrast, in 2D and 3D systems, CDWs occur due to *q*-dependent electron-phonon coupling, charge ordering, and PLD. However, the structural transition is common in all CDW materials [11]. The PLD leads to satellite reflections at distances $\pm q^{CDW}$ from the main Bragg reflections in single crystal X-ray diffraction studies (SXRD). The PLD introduces a new periodicity into the crystal and allows the analysis of CDW modulation [12]. In 3-D systems, first-order phase transition can be observed in physical properties, leading to incommensurate CDW. $EuAl_4$ [13], $SrAl_4$ [14], and $R_2Ir_3Si_5$ (where R= Ho, Er) [15,16] are examples of three-dimensional CDW compounds that show a thermal hysteresis in temperature-dependent electrical resistivity at the CDW transition temperature.

Compounds $R(Al_{1-x}Ga_x)_4$ (R = Sr, Eu, Ca) with the tetragonal $BaAl_4$ structure type with space group I4/mmm have recently received interest, because of the discoveries of tetragonal to orthorhombic CDW transitions in $EuAl_4$ and $SrAl_4$ [14], of a square skyrmion lattice phase in $EuAl_4$ [17–19], and of a tetragonal to monoclinic structural transition in $CaAl_4$ [20,21]. These materials show multiple antiferromagnetic transitions with topological hall effects. Eu-based compounds $Eu(Al_{1-x}Ga_x)_4$ reveal strong correlations between the structural, magnetic, and electronic properties [22,23]. These compounds contain three crystallographically independent atom sites where Ga preferably occupies one of the two independent Al sites **[Figure 1(a)]** [23]. $EuAl_4$ exhibits both the CDW at $T_{CDW}$= 145 K and orders antiferromagnetically below $T_{N1}$ = 15.4 K with three successive antiferromagnetic transitions at $T_{N2}$ = 13.2 K, $T_{N3}$ = 12.2 K and $T_{N4}$ = 10.0 K [24]. However, $EuGa_4$ orders antiferromagnetically below $T_N$= 16.4 K, but it develops a CDW only above a pressure of 2 GPa [24]. $EuAl_4$ contains skyrmionics spin texture; it only requires the alternate stacking of the magnetic Eu layer and nonmagnetic Al layer [25]. Generally, skyrmions appear in non-centrosymmetric materials, but magnetic skyrmions lattice structures are studied in other centrosymmetric compounds like MnNiGa [26], $Gd_2PdSi_3$ [27], $GdRu_2Si_2$ [28], $Gd_3Ru_4Al_{12}$ [29] including $EuAl_2Ga_2$ [22]. Skyrmions can be stabilized in these centrosymmetric compounds without a geometrically frustrated lattice. $BaNi_2As_2$ hosts incommensurate and commensurate CDWs at 142 K and 137 K, respectively, and shows symmetry breaking from tetragonal to orthorhombic at 142 K and then a first-order triclinic structural transition at 137 K [30,31]. A phase diagram for $BaNi_2(As_{1-x}P_x)_2$ shows that electronic nematic fluctuations induce a tetragonal-to-orthorhombic phase transition with the incommensurate charge density wave in the material [32]. The coexistence of CDW and superconductivity is well established in Kagome-lattice compounds $AV_3Sb_5$ (A= K,Rb,Cs) [33], whereas the coexistence of CDW and magnetic order is well established in FeGe [34,35].

The temperature dependence of electrical resistivity shows an anomaly in $EuAl_2Ga_2$, where the Al and Ga atoms split into two sublattices and create an ordered structure, that indicates a CDW state below 51 K [23]. The 3-dimensional CDW compound $EuAl_2Ga_2$ follows the RKKY interaction, having four antiferromagnetic transition temperatures: $T_{N1} \approx 19.5$ K, $T_{N2} \approx 15$ K, $T_{N3} \approx 11$ K, and $T_{N4} \approx 7$ K [22]. It also shows the persistence of the CDW in the magnetically ordered states [22]. The CDW modulation in $EuAl_4$ is orthorhombic with superspace group *Fmmm(00γ)s00* [18]. However, the atomic positions and lattice parameters of the basic structure of the CDW phase obey the tetragonal symmetry [18]. Recent studies on the CDW structure of $EuAl_4$ propose the loss of local inversion symmetry by convergent beam electron diffraction [36]. Alternatively, centrosymmetric *Immm(00γ)s00* superspace group symmetry was proposed for the CDW state of $EuAl_4$ based on inelastic x-ray scattering experiments, but this model doesn't fit SXRD data [37]. Resonant x-ray scattering (RXS) experiments on $EuGa_2Al_2$ have demonstrated the breaking of fourfold symmetry in CDW state that can be stabilized orthorhombic domains either *Immm(00γ)s00 or Fmmm(00γ)*s00 [38]. On the other hand, $SrAl_4$ shows an incommensurate CDW below $T_{CDW}$ = 243 K, which has the non-centrosymmetric orthorhombic *F222(00γ)*00s superspace group symmetry with modulation vector q = (0, 0, 0.1116(2)) at 200K [39].

A detailed structural description of CDW in $EuAl_2Ga_2$ has not yet been reported, and is presented here. We have found that CDW state of $EuAl_2Ga_2$ has an incommensurately modulated structure with orthorhombic superspace group *Immm(00γ)s*00, which differs from the symmetries of $EuAl_4$ and $SrAl_4$. The presence of CDW modulation below the magnetic transition temperature of $EuAl_2Ga_2$ is also observed.

**Experimental:**

### (A) Crystal Growth

Single crystals of $EuAl_2Ga_2$ have been grown by the Al-Ga self-flux method. The elements of europium, aluminum, and gallium were loaded into an alumina crucible in a ratio Eu:Al:Ga = 1:8:4 and subsequently sealed in a quartz-glass ampule under vacuum. The ampule was heated towards 1173 K, in order to obtain a homogenous solution. They were cooled at a rate of 2 K/hour to 873 K. At 873 K, the ampules were centrifuged to separate the single crystals from the flux. The stoichiometry of the crystals (i.e., Eu:Al:Ga = 1:2:2) has been confirmed using Energy-dispersive X-ray analysis (EDAX).

### (B) Magnetic Properties

The temperature dependence of the magnetic susceptibility of $EuAl_2Ga_2$ was measured on a single crystal using a commercial superconducting quantum interference device (SQUID) magnetometer (MPMS 7, Quantum Design, USA). Data were obtained after cooling the samples under zero-field cooling (ZFC) and field cooling (FC) conditions. A magnetic field of 1 kOe was applied along the [001] direction of the sample.

### (C) Single Crystal X-ray Diffraction (SXRD)

Temperature-dependent SXRD was carried out at beamline P24 of the synchrotron PETRA-III at DESY in Hamburg. A good-quality single crystal of $EuAl_2Ga_2$ was selected, with dimensions of 0.09mm × 0.055mm × 0.018mm. The SXRD was measured at a Eulerian diffractometer at temperatures down to 18 K using a CRYOCOOL open-flow helium gas cryostat. A Pilatus 1M- CdTe detector was used for data collection. Synchrotron radiation of wavelength 0.5000 Å was used for the experiment. Different runs employed different attenuators in the monochromatised beam, of filter factors of 1, 1.4804 and 10.5263. SXRD data sets were measured for temperatures of 298 K and 85 K (tetragonal phase) and of 35, 30 and 18 K (CDW phase). Each run of the data collection contained 3640 frames of $\Delta\varphi = 0.1$ deg and was repeated ten times. These data were binned to a data set of 364 frames of 1° of rotation and 10 sec exposure time per frame, using the programs *addrunscbf* and *combcbf* [40]. A single run at 298 K was recorded at 110 mm crystal-to-detector distance. Two runs were needed at 35, 30, and 18 K, with and without a 2θ offset, for a 260 mm crystal to detector distance. An improved spatial resolution was achieved at 35, 30 and 18 K through a crystal-to-detector distance of 270 mm. A similar resolution of the SXRD data then required runs with offsets of 0 and 25 deg in 2θ.

CrysAlis Pro [41] makes an unwarp image that is based on all of the measured scattering, including scattering between the Bragg reflections (**Figure 2)**. The EVAL15 [42] software suite was used for data processing of the binned SXRD data. The two runs with different offsets in 2θ were integrated separately for each of the data measured at 35K, 30 K and 18 K. They were combined by the module ANY of EVAL15. An outlier rejection was based on tetragonal *4/mmm* and orthorhombic *mmm* Laue classes. Absorption correction and scaling were performed in *SADABS* [43] with Laue symmetries *4/mmm* for all five data sets and *mmm* for the data measured at 35, 30 and 18 K.

The resulting reflection files were imported into JANA2020 [44] for structure solution and refinement. The Laue symmetry *4/mmm* and *mmm* is used for the periodic and CDW phases of $EuAl_2Ga_2$. We have used the superspace approach to index and integrate the incommensurately modulated data to solve the crystal structure in the CDW phase [12,45]. The crystal structure is drawn using the program VESTA [46].

**Results and discussion:**

### (A) Periodic Structure and Magnetic Properties

SXRD data at 298 K and 85 K confirmed the $BaAl_4$ structure type with space group *I4/mmm* [Table 1, **Figure 1(a)]**. In this structure, the $Eu^{2+}$ ions form a square layer in the ab plane. Neighboring Eu layers are translated by (0.5, 0.5, 0.5) due to the *I*-centering. The Al and Ga atoms lie between neighboring Eu layers and are ordered on Wyckoff positions 4d and 4e, respectively. The refined composition ($EuAl_{2.023\,(21)}Ga_{1.977}$) is close to the nominal composition $EuAl_2Ga_2$. Since Ga thus substitutes for the Al2 atom of $EuAl_4$, the atomic sites in $EuAl_2Ga_2$ are designated as Eu1, Al1, and Ga2. The substitution of Al by Ga results in a reduced volume of $EuAl_2Ga_2$ as compared to $EuAl_4$ [18] and $SrAl_4$ [39].

Previous studies have found that Ga-doped $Eu(Al_{1-x}Ga_x)_4$ compounds exhibit antiferromagnetic transitions below 25 K [23]. **Figure 1(b-d)** shows the temperature-dependent zero-field cooled (ZFC) and field cooled (FC) *dc* magnetization of $EuAl_2Ga_2$ in the temperature range of 2 K – 100 K, and the presence of 1k Oe magnetic field applied parallel to the c-axis. The magnetization curve shows an antiferromagnetic nature of the sample where the FC magnetization follows the trend of ZFC magnetization. The inverse susceptibility curve obtained from the ZFC magnetization is shown in **Figure 1(c)**. A Curie-Weiss fit to the paramagnetic region (55–100 K) resulted in a positive Curie-Weiss temperature that indicates dominant ferromagnetic interactions, despite the antiferromagnetic ordering at lower temperatures. The effective magnetic moment ($\mu_{eff}$) for $EuAl_2Ga_2$ is 7.917 $\mu_B$, which is close to the theoretical of 7.937 $\mu_B$.

Neutron diffraction studies have provided the incommensurate helical and cycloidal magnetic phases of antiferromagnetic $EuAl_2Ga_2$ below $T_N$ [22]. Multiple magnetic transitions can be identified by the dips in the first derivative of magnetization (dM/dT) curve. These transition temperatures are $T_{N1}$ = 21.4 K, $T_{N2}$ = 15.78 K, $T_{N3}$ = 11.02 K, and $T_{N4}$ = 7.45 K, for a magnetic field of 1 kOe applied parallel to the **c** axis, as shown in **Figure 1(d)**. These transition temperatures are very close to those reported by *Stavinoha et al.* [23].

### (B) Temperature-dependent CDW structure

The first-order, CDW phase transition of $EuAl_2Ga_2$ has been reported to occur at $T_{CDW}$ ≈ 51 K experiments [22,23]. **Figure 2** shows parts of (*h* 0 *l*) sections of reciprocal space, reconstructed from SXRD data at four different temperatures. First-order satellites are clearly visible along **c*** at 35 K, 30 K and 18 K. They represent the incommensurately modulated CDW with wave vectors (0, 0, 0.1141(2)) at 35 K, (0, 0, 0.1139 (2)) at 30 K and (0, 0, 0.1126 (4)) at 18 K. The value of the incommensurate component decreases with decreasing temperature. Higher-order satellites have not been observed.

We have observed that $EuAl_2Ga_2$ preserves tetragonal *I4/mmm* symmetry as symmetry of the basic structure in the CDW state, as it was also found for $EuAl_4$ and $SrAl_4$ [18, 40]. The incommensurate CDW phase of $EuAl_4$ has the orthorhombic symmetry *Fmmm(00γ)s00* with modulation wave vector **q** = (0, 0, 0.1781(3)) at 70 K and **q** = (0 0 0.1741(2)) at 20 K [18]. For the CDW phase of $EuAl_2Ga_2$, the present SXRD data indicate symmetry according to the superspace group *Immm(00γ)s00*, as detailed below. This finding is in line with the results of Resonant Inelastic X-ray scattering (RIXS), which indicated the loss of the fourfold symmetry in in the CDW state of $EuAl_2Ga_2$ [38].

In order to establish the crystal structure and symmetry of the CDW state of $EuAl_2Ga_2$, the SXRD data were initially processed according to 4/mmm point symmetry. Averaging in JANA2020 showed that these SXRD data fulfil 4/mmm point symmetry very well according to the observed values of $R_{int}$ (see **Table 2**). However, structure refinements at 35 K, 30 K and 18 K employing a tetragonal superspace group lead to high values for $R_F$ (satellite) (Table 2 and Tables S4 and S5). Exception is the superspace group *I422(00γ)q00*, but R values attained with this non-centrosymmetric superspace group are still significantly higher than those for the orthorhombic choices. Therefore, these results confirm that the CDW phase doesn't have tetragonal symmetry. Secondly, structure models were employed with superspace groups based on orthorhombic subgroups of I4/mmm. SXRD data processed in 4/mmm symmetry imply two orthorhombic domains with twin fractions 0.5 : 0.5. Structure refinements then led to comparable R values for orthorhombic superspace groups *Fmmm(00γ)s00* and *Immm(00γ)s00* **(Table 2).**

In a second approach, structure refinements of models with *Fmmm(00γ)s00* and *Immm(00γ)s00* symmetries have been performed against the SXRD data processed according to the respective orthorhombic point symmetries *Fmmm* and *Immm*. Here, *R* values indicate a better fit for *Immm(00γ)s00* than for *Fmmm(00γ)s00* for all the temperatures (**Table 2)**. Volumes for the twin domains were refined and they deviate significantly from 0.5 : 0.5 for the structure in *Immm(00γ)s00* (**Table 1**). **Table 3** provides the basic-structure coordinates of $EuAl_2Ga_2$ at 298 K and 85 K with *I4/mmm* symmetry, and at 35 K, 30 K and 18 K with orthorhombic *Immm(00γ)s00* symmetry. The basic structure of $EuAl_2Ga_2$ retains tetragonal symmetry at all temperatures, since the z-coordinate of Al1 does not deviate significantly from 0.25. Details on the structure model with *Fmmm(00γ)s00* symmetry are proved in **Tables S1–S3** in the Supplementary Material.

Recently, we have found for $SrAl_4$ that a non-centrosymmetric structure with symmetry *F222(00γ)00s* give a better fit to the SXRD data than its supergroup *Fmmm(00γ)s00* [36,39]. Main argument was a better fit to the second-order satellites for the non-centrosymmetric model [39]. For $EuAl_4$, the SXRD data set did not contain second-order satellites, and the centrosymmetric and non-centrosymmetric models could not be distinguished [18,39]. Presently, without second-order satellites for $EuAl_2Ga_2$, non-centrosymmetric superspace groups do lead to better fits to the data than the superspace group *Immm(00γ)s00* (**Table 2**). Furthermore, refinements are unstable for superspace groups based on point symmetry *222*, thus indicating dependencies between the parameters. A complete list of non-centrosymmetric space groups for 18 K is shown in **Table S4** of Supplementary Material. In conclusion, the present SXRD data lead to a centrosymmetric structure model for the CDW state of $EuAl_2Ga_2$, which has orthorhombic symmetry *Immm(00γ)s00*.

### (C) Thermal expansion

The present data reveal negative thermal expansion (NTE) below the CDW transition (**Table 1; Figure 3**). It has been shown that NTE can be enhanced by substituting Al, Ga, and Cr in various intermetallics [47], which also enhances the magnetic transition temperature [48]. This general feature thus explains the present observation of NTE of $EuAl_2Ga_2$. NTE has been found in other $RT_2X_2$ compounds like $YbMn_2Ge_2$ compound over a wide temperature range [49]. The NTE ($\alpha_V = \frac{1}{V}(\frac{dV}{dT})$) at the temperature difference of 35K-30K and 30K-18K are -2.06 × $10^{-3}$ $K^{-1}$ and -2.17× $10^{-4}$ $K^{-1}$ observed with space group *Immm(00γ)s00.*

### (D) Location of the CDW

Structural parameters, such as atomic displacements, atomic distances and bond angles, can be evaluated as a function of the phase *t* of the modulation wave. Such *t*-plots are useful for analyzing the correlated variations of different parameters, e.g. of the atomic environments. The displacement modulation is described by a modulation function for each atom of the form [12],

$$u_{\alpha}(\bar{x}_{s4}) = \sum_{n}\{A_{\alpha,n}\sin(2n\pi\bar{x}_{s4}) + B_{\alpha,n}\cos(2n\pi\bar{x}_{s4})\} \quad (1)$$

where n = 1 due to first order harmonics and α = x, y, z. The fourth superspace coordinate for the $j^{\text{th}}$ atom can be represented as, $\bar{x}_{s4}(j) = t + \boldsymbol{q}.\bar{\boldsymbol{x}}(j)$ where $t$ is the initial phase of modulation and $\bar{\boldsymbol{x}}(j)$ is the position of the $j^{\text{th}}$ atom in the periodic basic structure [12]. The observed amplitudes for displacement modulation show that atoms have displacements ($u_x$) along the *x* direction and zero displacements $u_y$ along y and $u_z$ along z for the structure models with superspace group *Immm(00γ)s00* **[Table 4** and **Figure 4 (a-c)]**. The non-zero modulation amplitudes for all three atoms contribute to CDW order in the material. However, the Eu1 atom is elastically coupled to Al1 and Ga2 atoms, having the *sine* component of modulation function, whereas Al1 and Ga2 have both periodic components of *sine* and *cosine* waves along the *x* direction. Displacement modulation for all the atoms is a transverse wave.

To understand how Al1 and Ga2 atoms contribute to CDW ordering in the material, we consider the shortest interatomic distances, which are Ga2–Ga2, Ga2–Al1 and Al1–Al1 distances. *t*-plots of these interatomic distances show that the largest modulation is found for the Al1–Al1 distances at all three temperatures (**Figures 5 and 6; Figures S1–S3)**. This finding suggests that the CDW resides on the Al atoms, like for $EuAl_4$ [18]. The different symmetries of the CDWs in $EuAl_2Ga_2$ and $EuAl_4$ [18] can be attributed to the presence of Ga in $EuAl_2Ga_2$.

## Conclusions

Like $EuGa_4$, $EuAl_4$ and $SrAl_4$; $EuAl_2Ga_2$ has the $BaAl_4$ structure type with tetragonal symmetry I4/mmm at room temperature. SXRD shows the presence of satellite reflections along **c*** at temperatures below $T_{CDW}$ = 51 K. The incommensurately modulated CDW crystal structure of $EuAl_2Ga_2$ was found to possess symmetry according to the superspace group *Immm(00γ)s00*. This is different from $EuAl_4$ that has superspace symmetry *Fmmm(00γ)s00* for its CDW state, and it is different from $EuGa_4$, which does not develop a CDW at ambient pressure. The different symmetries of the CDW states of $EuAl_4$ and $EuAl_2Ga_2$ might thus be explained by the effects of Ga substitution in the latter compound. In this respect, it should be noted that both structure models, *Fmmm(00γ)s00* and *Immm(00γ)s00*, lead to a large variation of the Al1–Al1 contact distance as principal effect of the modulation. Either model thus explains the presence of the CDW on the Al1 atoms.

We have found negative thermal expansion (NTE) below $T_{CDW}$. The NTE effect can be rationalized as the result of Ga substitution into this intermetallic material [45].

**Acknowledgement:**

We thank Kerstin Küspert and Franz Fischer for technical assistance with the experiments. We acknowledge DESY (Hamburg, Germany), a member of the Helmholtz Association HGF, for the provision of experimental facilities. Parts of this research were carried out at PETRA-III, and we would like to thank Heiko Schulz-Ritter for assistance in using beamline P24. Beamtime was allocated for proposal I-20211203.

**Table 1**. Crystallographic data of $EuAl_2Ga_2$ at temperatures of 298 K, 85 K, 35 K, 30 K and 18 K. The criterion of observability is $I > 3\ \sigma(I)$. a, b, c are the lattice parameters, V is the volume of the unit cell of the basic structure, and $q_z$ is the z component of the modulation wave vector. D is the detector distance; twvol is twin volume.

| Temperature | 298 K | 85 K | 35 K | 30 K | 18 K |
|---|---|---|---|---|---|
| Crystal System | Tetragonal | Tetragonal | Orthorhombic | Orthorhombic | Orthorhombic |
| Space Group | *I4/mmm* | *I4/mmm* | *Immm(00γ)s00* | *Immm(00γ)s00* | *Immm(00γ)s00* |
| No. | 139 | 139 | 71.1.12.2 | 71.1.12.2 | 71.1.12.2 |
| Laue symmetry | 4/*mmm* | 4/*mmm* | *mmm* | *mmm* | *mmm* |
| a (Å) | 4.3389(38) | 4.33410(16) | 4.29366(14) | 4.30768(14) | 4.31351(25) |
| b (Å) | 4.3389 | 4.33410 | 4.29677(10) | 4.31402(13) | 4.31411(19) |
| c (Å) | 10.9674(77) | 10.95703(47) | 10.86970(25) | 10.90222(31) | 10.91571(54) |
| V ($Å^3$) | 206.47(41) | 205.822(20) | 200.534(9) | 202.601(12) | 203.130(20) |
| $q_z$ | - | - | 0.1142(2) | 0.1139 (2) | 0.1126(4) |
| D (mm) | 110 | 260 | 260 | 260 | 260 |
| Filter factor | 10.5263 | 10.5263 | 1.4804+10.5263 | 0+10.5263 | 0+10.5263 |
| 2θ offset | 0 | 0, 25 | 0, 25 | 0, 25 | 0, 25 |
| $[\sin(\theta)/\lambda]_{max}$ $Å^{-1}$ | 0.690825 | 0.730126 | 0.734924 | 0.732604 | 0.731724 |
| No. of Reflections | | | | | |
| Main (obs/all) | 105/108 | 119/122 | 153/174 | 142/161 | 148/168 |
| Satellites (obs/all) | - | - | 179/333 | 140/331 | 150/332 |
| $R_{int}$ (obs/all) % | 3.67/3.67 | 1.78/1.78 | 2.97/3.03 | 2.33/2.41 | 2.17/2.22 |
| No. of parameters | 9 | 9 | 16 | 16 | 16 |
| $R_F$ (all) % | 1.53 | 1.60 | 3.47 | 3.23 | 2.50 |
| $_wR_F^2$ (all)% | 1.99 | 2.12 | 4.67 | 3.74 | 2.85 |
| $R_F$ (main) % | 1.53 | 1.59 | 2.56 | 2.63 | 2.07 |
| $R_F$ (sat) % | - | - | 7.84 | 6.44 | 4.70 |
| GOF (obs/all) | 1.64/1.61 | 1.66/1.64 | 1.71/1.41 | 1.31/1.01 | 0.98/0.77 |
| twvol1/twvol2 | - | - | 0.386(8)/0.614 | 0.419(9)/0.581 | 0.447(6)/0.552 |

**Table 2**. R values (%) of different structure models on the basis of different centrosymmetric and non-centrosymmetric superspace groups for SXRD data at 35 K, 30 K, 18 K.

| Superspace group | averaging | $R_{int}$ | $R_F$ (all) | $R_F$ (Main) | $R_F$ (Satellite) | $_wR_F^2$ (all) | Number of Reflections (Main / Satellite) | Parameter |
|---|---|---|---|---|---|---|---|---|
| Temperature = 35 K | | | | | | | | |
| *I4/mmm(00γ)000* | 4*/mmm* | 3.28 | 12.73 | 3.12 | 59.91 | 19.92 | 99/105 | 13 |
| *Fmmm(00γ)s00* | 4*/mmm* | 3.28 | 3.02 | 1.91 | 8.46 | 3.56 | 99/105 | 14 |
| *Immm(00γ)s00* | 4*/mmm* | 3.28 | 2.98 | 1.90 | 8.30 | 3.58 | 99/105 | 15 |
| *Fmmm(00γ)s00* | *mmm* | 2.77 | 3.59 | 2.40 | 9.57 | 4.39 | 147/161 | 15 |
| *F222(00γ)00s* | *mmm* | 2.77 | 3.55 | 2.39 | 9.38 | 4.36 | 147/161 | 19 |
| *Immm(00γ)s00* | *mmm* | 2.97 | 3.47 | 2.56 | 7.84 | 4.67 | 153/179 | 16 |
| *I222(00γ)00s* | *mmm* | 2.97 | 3.46 | 2.56 | 7.76 | 4.66 | 153/179 | 20 |
| *I422(00γ)q00* | *mmm* | 2.97 | 3.82 | 2.67 | 9.36 | 5.47 | 153/179 | 14 |
| *Temperature = 30 K* | | | | | | | | |
| *I4/mmm(00γ)000* | 4*/mmm* | 2.88 | 15.31 | 4.75 | 64.98 | 23.98 | 92/105 | 13 |
| *Fmmm(00γ)s00* | 4*/mmm* | 2.88 | 2.77 | 1.96 | 6.53 | 3.31 | 92/105 | 14 |
| *Immm(00γ)s00* | 4*/mmm* | 2.88 | 2.75 | 1.97 | 6.43 | 3.35 | 92/105 | 15 |
| *Fmmm(00γ)s00* | *mmm* | 2.89 | 3.82 | 2.82 | 8.60 | 4.66 | 148/162 | 15 |
| *F222(00γ)00s* | *mmm* | 2.89 | 3.76 | 2.80 | 8.46 | 4.63 | 148/162 | 19 |
| *Immm(00γ)s00* | *mmm* | 2.33 | 3.24 | 2.63 | 6.45 | 3.74 | 142/140 | 16 |
| *I222(00γ)00s* | *mmm* | 2.33 | 3.22 | 2.63 | 6.37 | 3.74 | 142/140 | 20 |
| *I422(00γ)q00* | *mmm* | 2.33 | 3.39 | 2.64 | 7.39 | 3.99 | 142/140 | 14 |
| *Temperature = 18 K* | | | | | | | | |
| *I4/mmm(00γ)000* | 4*/mmm* | 2.69 | 12.66 | 2.88 | 59.68 | 19.70 | 97/105 | 13 |
| *Fmmm(00γ)s00* | 4*/mmm* | 2.69 | 2.13 | 1.56 | 4.85 | 2.45 | 97/105 | 14 |
| *Immm(00γ)s00* | 4*/mmm* | 2.69 | 2.13 | 1.56 | 4.85 | 2.47 | 97/105 | 15 |
| *Fmmm(00γ)s00* | *mmm* | 2.60 | 2.91 | 2.27 | 6.12 | 3.31 | 151/150 | 15 |
| *F222(00γ)00s* | *mmm* | 2.60 | 2.87 | 2.27 | 5.92 | 3.28 | 151/150 | 19 |
| *Immm(00γ)s00* | *mmm* | 2.17 | 2.50 | 2.07 | 4.71 | 2.86 | 148/150 | 16 |
| *I222(00γ)00s* | *mmm* | 2.17 | 2.49 | 2.07 | 4.64 | 2.85 | 148/150 | 20 |
| *I422(00g)q00* | *mmm* | 2.17 | 2.59 | 2.11 | 5.05 | 3.01 | 148/150 | 14 |

**Table 3.** Atomic coordinates of $EuAl_2Ga_2$ as obtained from structure refinements against SXRD data at 298 K and 85 K (crystal structure), and at 35 K, 30 K, and 18 K (basic structure).

| Atoms | x | y | z | U11 | U22 | U33 | U12 | U13 | U23 | Ueq |
|---|---|---|---|---|---|---|---|---|---|---|
| 298 K (*I4/mmm*) | | | | | | | | | | |
| Eu1 | 0 | 0 | 0 | 0.0087(3) | 0.0087(3) | 0.0096(3) | 0 | 0 | 0 | 0.0090(1) |
| Al1 | 0 | 0.5 | 0.25 | 0.0085(5) | 0.0085(5) | 0.0067(7) | 0 | 0 | 0 | 0.0079(3) |
| Ga2 | 0 | 0 | 0.38561(6) | 0.0112(3) | 0.0112(3) | 0.0073(4) | 0 | 0 | 0 | 0.0073(4) |
| 85 K (*I4/mmm*) | | | | | | | | | | |
| Eu1 | 0 | 0 | 0 | 0.0030(2) | 0.0030(2) | 0.0039(3) | 0 | 0 | 0 | 0.0033(1) |
| Al1 | 0 | 0.5 | 0.25 | 0.0030(5) | 0.0030(5) | 0.0031(7) | 0 | 0 | 0 | 0.0030(3) |
| Ga2 | 0 | 0 | 0.38564(6) | 0.0045(3) | 0.0045(3) | 0.0041(3) | 0 | 0 | 0 | 0.0043(1) |
| 35 K (*Immm(00γ)s00*) | | | | | | | | | | |
| Eu1 | 0 | 0 | 0 | 0.0032(3) | 0.0032(3) | 0.0034(4) | 0 | 0 | 0 | 0.0033(2) |
| Al1 | 0 | 0.5 | 0.2514(8) | 0.0030(10) | 0.0030(10) | 0.0011(14) | 0 | 0 | 0 | 0.0024(5) |
| Ga2 | 0 | 0 | 0.38566(9) | 0.0040(4) | 0.0040(4) | 0.0031(5) | 0 | 0 | 0 | 0.0037(2) |
| 30 K (*Immm(00γ)s00*) | | | | | | | | | | |
| Eu1 | 0 | 0 | 0 | 0.0047(3) | 0.0047(3) | 0.0037(3) | 0 | 0 | 0 | 0.00433(17) |
| Al1 | 0 | 0.5 | 0.2497(10) | 0.0025(9) | 0.0025(9) | 0.0010(12) | 0 | 0 | 0 | 0.0020(6) |
| Ga2 | 0 | 0 | 0.38562(8) | 0.0059(4) | 0.0059(4) | 0.0032(4) | 0 | 0 | 0 | 0.0050(2) |
| 18 K (*Immm(00γ)s00*) | | | | | | | | | | |
| Eu1 | 0 | 0 | 0 | 0.0031(2) | 0.0031(2) | 0.0019(2) | 0 | 0 | 0 | 0.00271(12) |
| Al1 | 0 | 0.5 | 0.2486(7) | 0.0025(6) | 0.0025(6) | 0.0001(8) | 0 | 0 | 0 | 0.0017(4) |
| Ga2 | 0 | 0 | 0.38565(6) | 0.0045(3) | 0.0045(3) | 0.0017(3) | 0 | 0 | 0 | 0.00358(16) |

**Table 4.** Amplitudes of the modulation functions of structure models for $EuAl_2Ga_2$ at 35 K, 30 K and 18 K, for the superspace group *Immm(00γ)s00*.

| Atoms | $A_{1,x}$ a (Å) | $A_{1,y}$ b (Å) | $A_{1,z}$ c (Å) | $B_{1,x}$ a (Å) | $B_{1,y}$ b (Å) | $B_{1,z}$ c (Å) |
|---|---|---|---|---|---|---|
| 35 K (*Immm(00γ)s00*) | | | | | | |
| Eu1 | 0.07979(35) | 0 | 0 | 0 | 0 | 0 |
| Al1 | 0.07629(133) | 0 | 0 | -0.02742(155) | 0 | 0 |
| Ga2 | 0.07999(56) | 0 | 0 | 0.02860(45) | 0 | 0 |
| 30 K (*Immm(00γ)s00*) | | | | | | |
| Eu1 | 0.07804(38) | 0 | 0 | 0 | 0 | 0 |
| Al1 | 0.07420(139) | 0 | 0 | -0.02489(170) | 0 | 0 |
| Ga2 | 0.07968(61) | 0 | 0 | 0.02864(51) | 0 | 0 |
| 18 K (*Immm(00γ)s00*) | | | | | | |
| Eu1 | 0.07987(28) | 0 | 0 | 0 | 0 | 0 |
| Al1 | 0.07869(115) | 0 | 0 | -0.02435(132) | 0 | 0 |
| Ga2 | 0.07975(45) | 0 | 0 | 0.02813(39) | 0 | 0 |

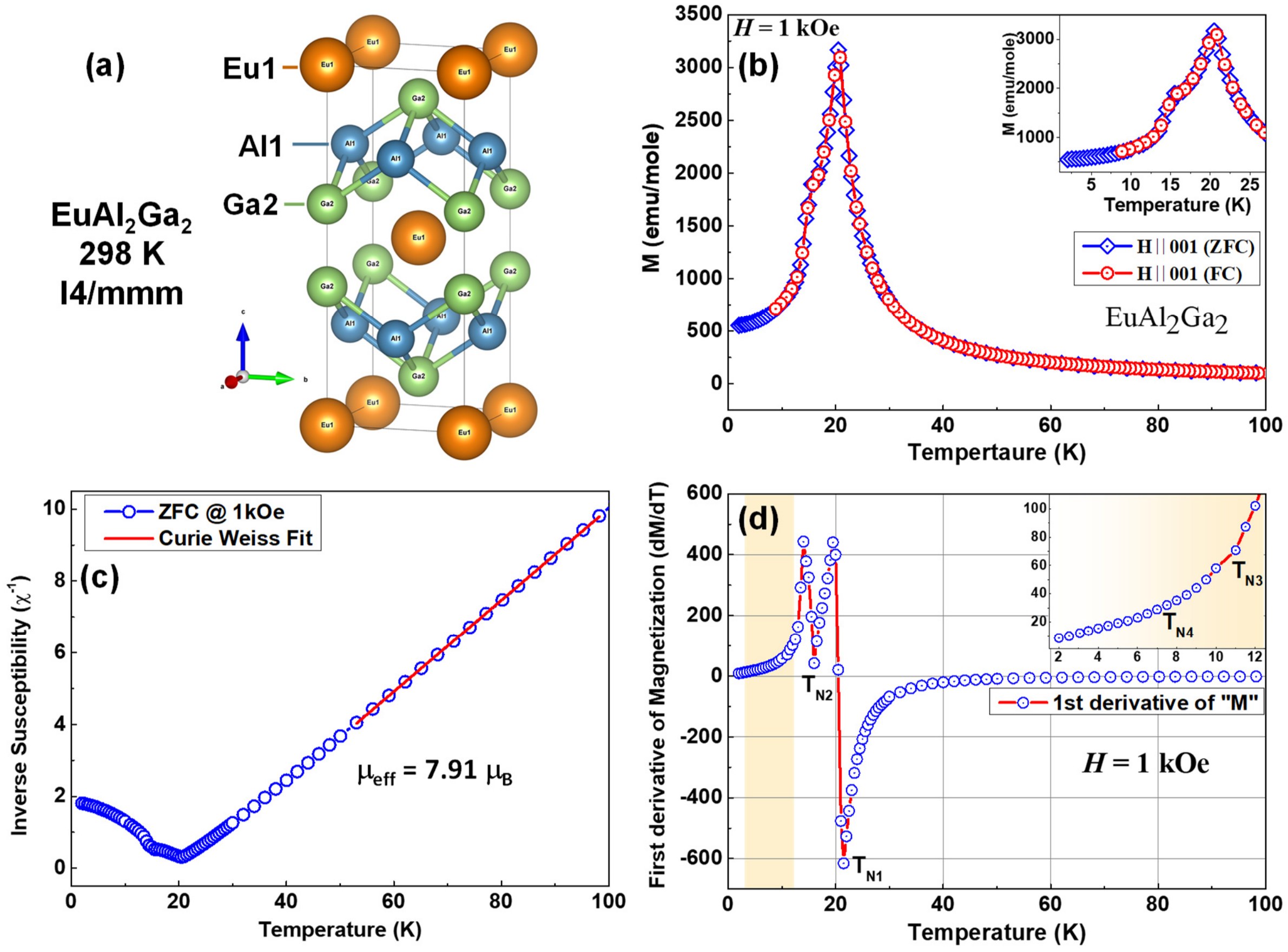

Figure 1: (a) Tetragonal structure of $EuAl_2Ga_2$ at 298K having the centrosymmetric *I*4/*mmm* space group. (b) Temperature dependent zero field cooled, field cooled dc magnetization observed at 1k Oe in the temperature range of 2K – 100 K, showing the antiferromagnetic nature due to $Eu^{2+}$ ion; (c) Inverse susceptibility obtained by ZFC curve; a Curie-Weiss model was fitted to the data between 55 and 100 K (red line). (d) first derivative of magnetization showing the transition temperatures of the material at $T_{N1} \approx 21.4$ K, $T_{N2} \approx 15.78$ K, $T_{N3} \approx 11.02$ K, $T_{N4} \approx 7.45$ K.

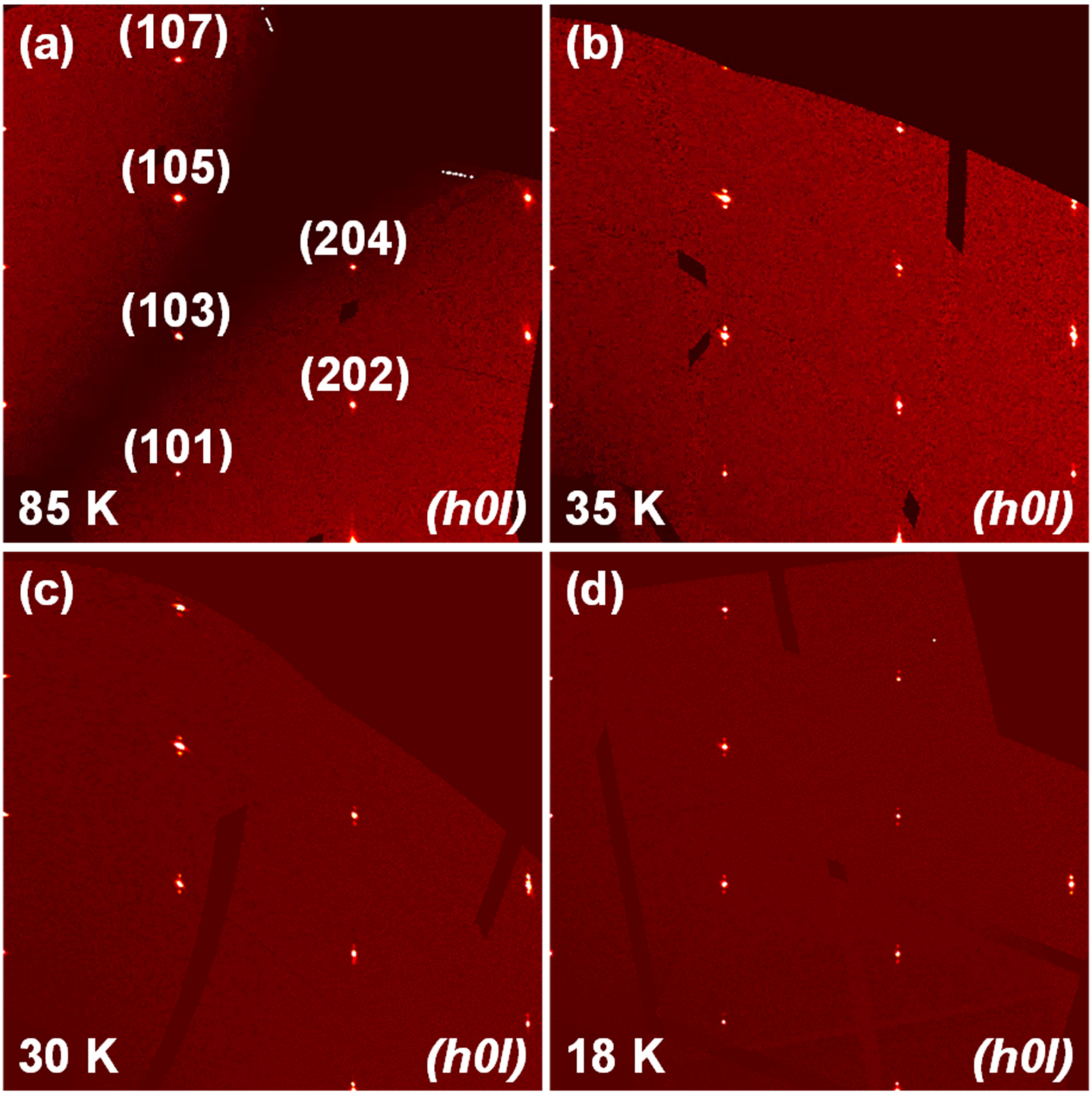

Figure 2: *(h 0 l)* Sections of reciprocal space reconstructed from SXRD data measured on $EuAl_2Ga_2$ at (a) T = 85 K, (b) T = 35 K, (c) 30 K, and (d) 18 K. Dark areas are due to insensitive parts of the detector.

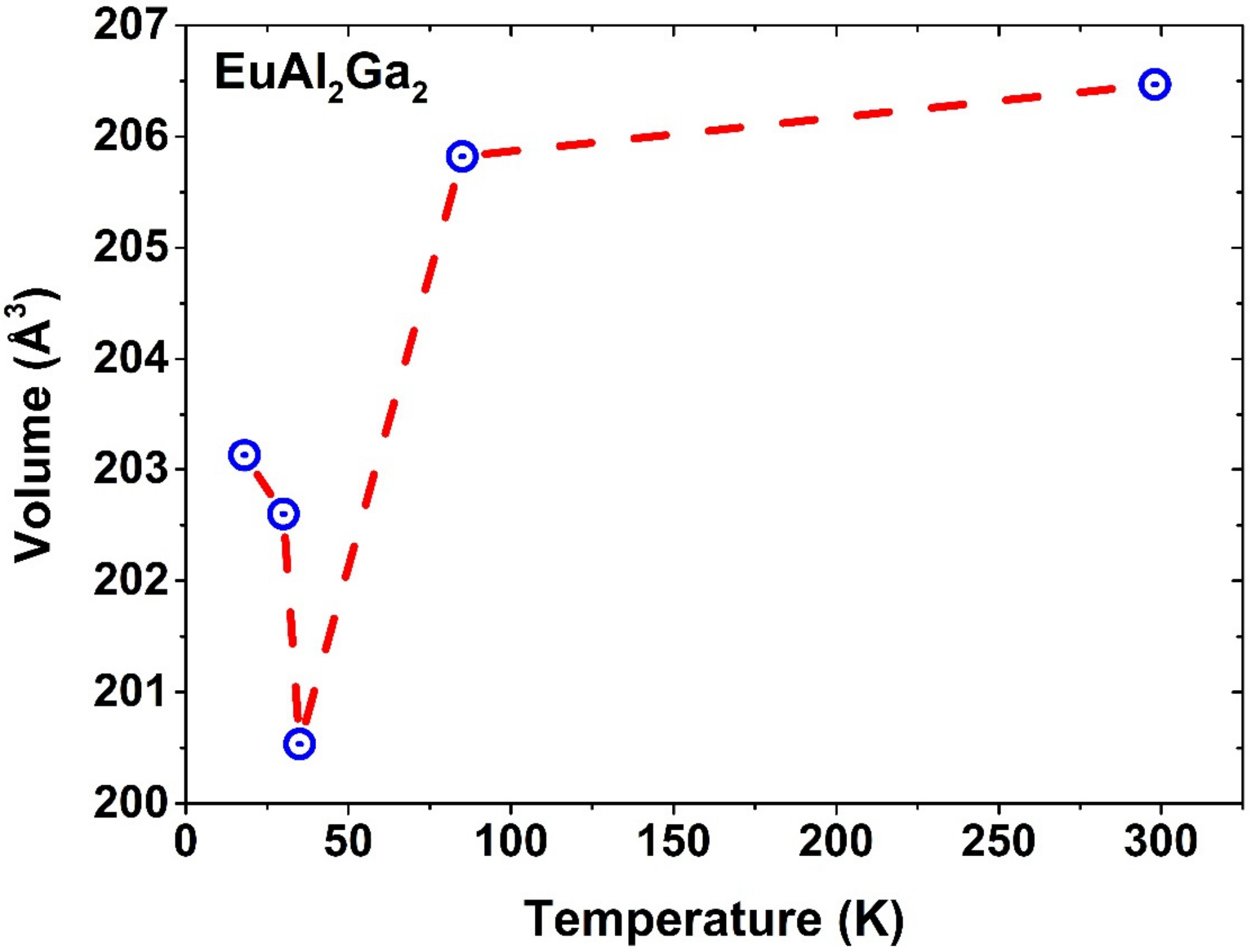

Figure 3: Temperature dependence of the volume of the unit cell.

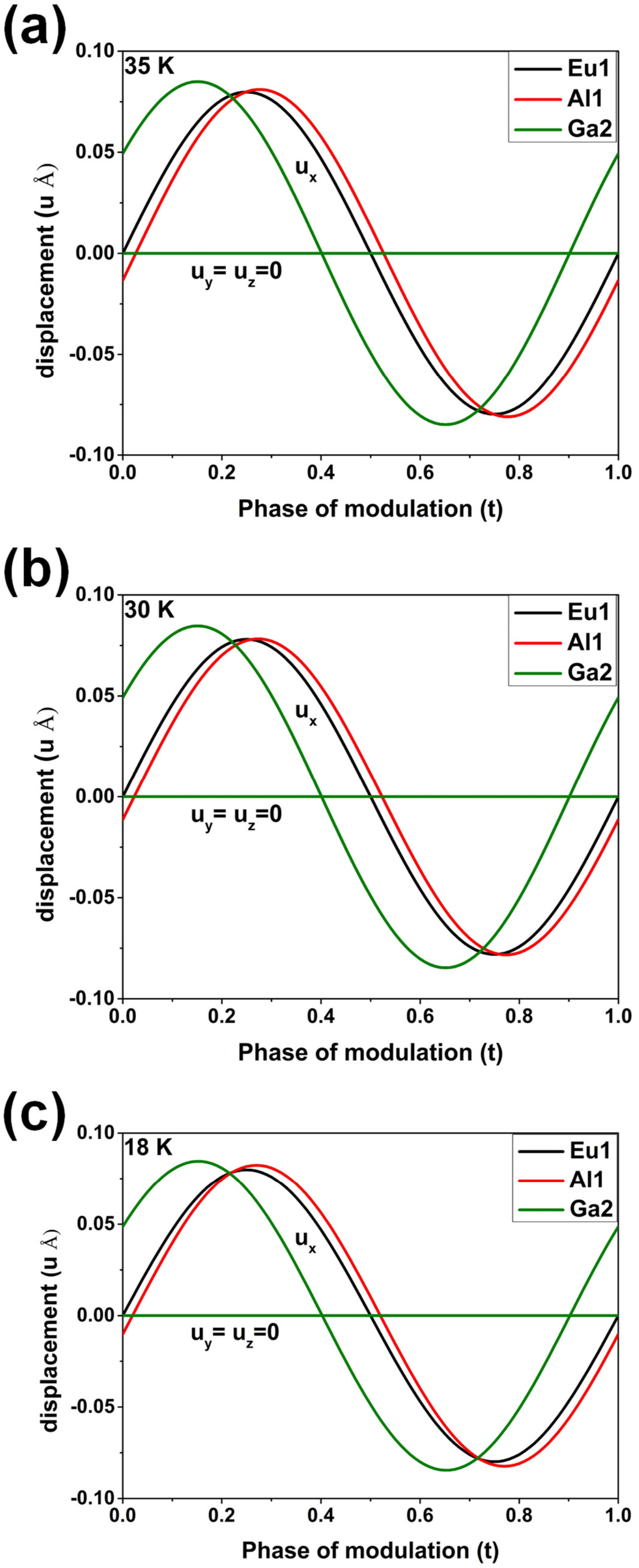

Figure 4: t-plot of the modulation functions for the displacement of atoms Eu1, Al1 and Ga2 at 35 K, 30K, 18K for superspace group *Immm(00γ)s00*.

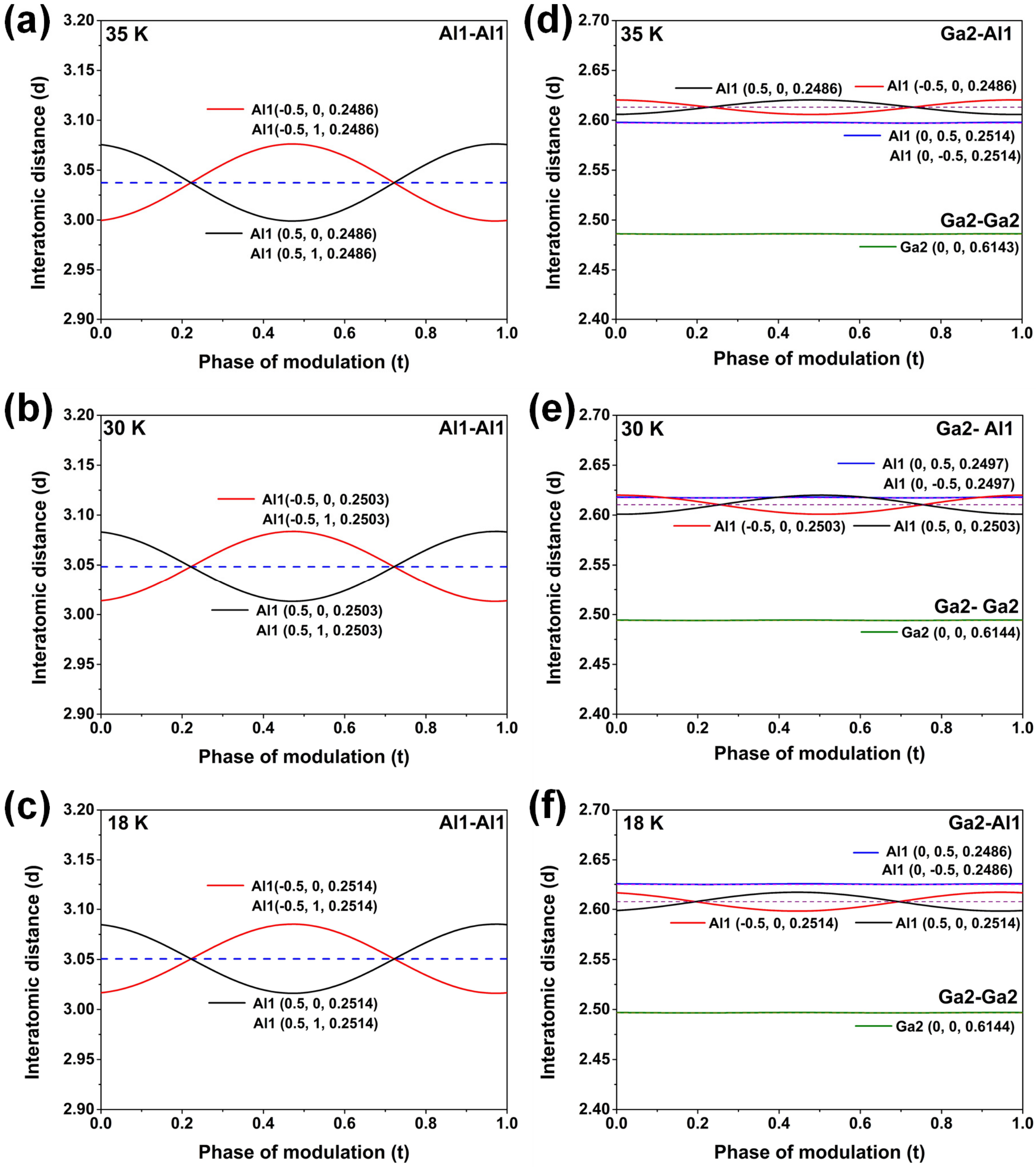

Figure 5: t-plot of interatomic distances (Å) d[Al1-Al1], d[Ga2-Al1], d[Ga2-Ga2] at 35K, 30 K and 18 K for superspace group *Immm(00γ)s00*. Compare to Figure 6.

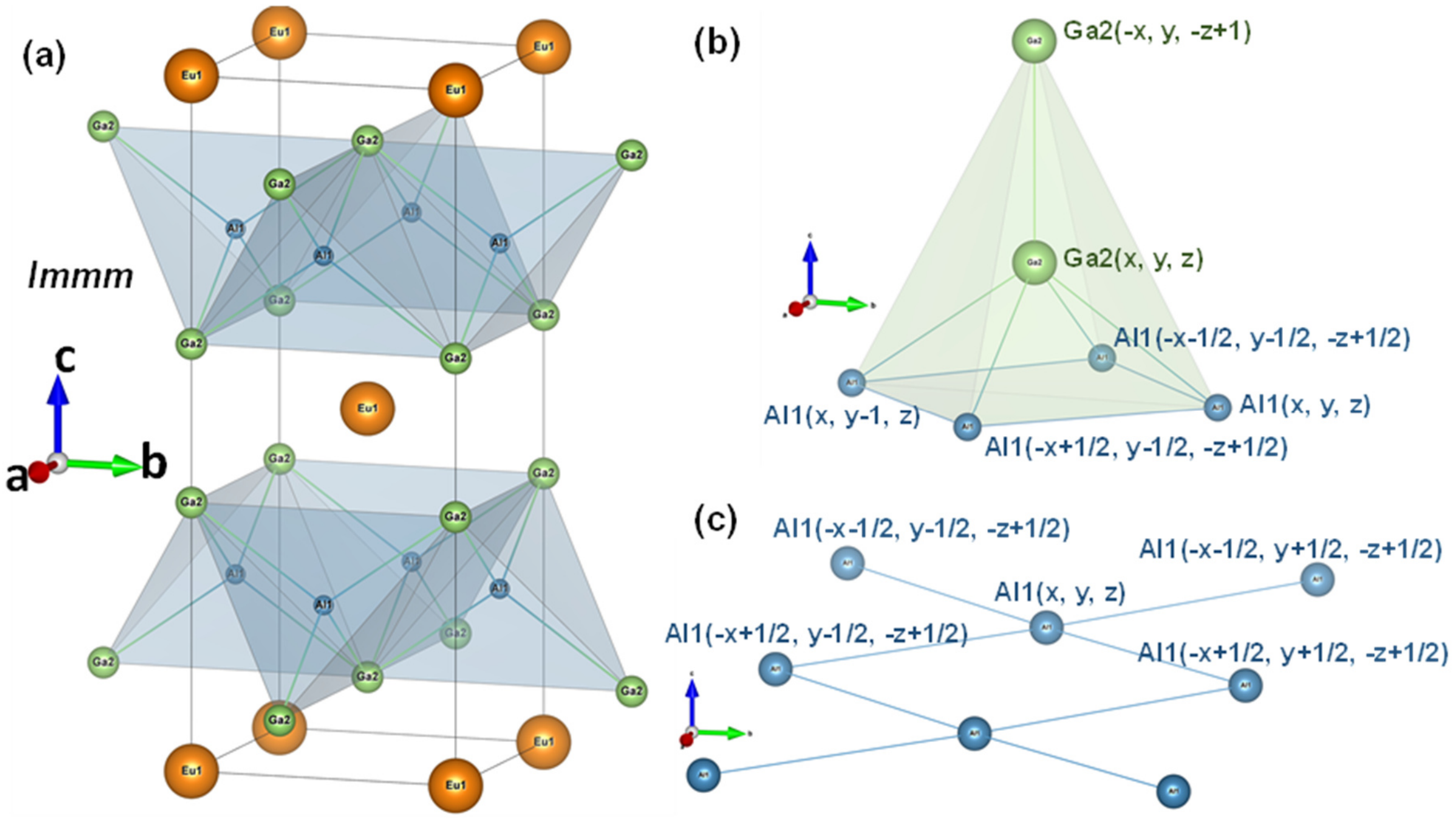

Figure 6: Crystal Structure of $EuAl_2Ga_2$ having orthorhombic structure with *Immm* symmetry at 18 K (a) Polyhedral view of Al1-Ga2 atoms; (b) Neighbouring atoms of Ga2 for which the CDW modes between Ga2-Ga2 and Ga2-Al1 atoms are shown in figure 5; (c) neighbouring Al atoms of Al1 for which the Al1-Al1 layers shows largest modulation of interatomic distances causing CDW in the material.